\documentclass[floatfix,aps,prl,reprint,groupedaddress,10pt]{revtex4-1}

\usepackage{graphicx,amsmath}

\begin{document}

\title{Kerr breather-soliton time crystals}

\author{Daniel C. Cole}
\email[]{daniel.cole@nist.gov}

\author{Scott B. Papp}
\email{scott.papp@nist.gov}
\affiliation{Time \& Frequency Division, NIST Boulder\\Boulder, Colorado 80305, USA}

\date{\today}

\begin{abstract}
	Dissipative solitons in passive, driven, Kerr-nonlinear optical resonators are circulating pulses of light that are sustained by a pump laser and can propagate indefinitely. These solitons are of interest both from the standpoint of fundamental nonlinear dynamics and because they could bring optical frequency synthesis and metrology capabilities to the chip scale. Under some conditions Kerr solitons can exhibit breathing behavior, in which case their amplitude, bandwidth, and duration oscillate in time over the course of many resonator round trips. These breather oscillations emerge spontaneously, and their frequency is determined by the system parameters. Here we propose and theoretically investigate the possibility that these breather oscillations may become subharmonically entrained to the periodic perturbation occurring at the resonator round-trip time $T_R$ that arises from out-coupling and pumping. The result of subharmonic entrainment is an integer ratio $T_b/T_R=N\gg1$ between the breather period $T_b$ and the round-trip time. We find that in a regime of intermediate resonator finesse ($\mathcal{F}\sim30-40$), rigid subharmonic entrainment---or locking of the breather oscillation to the round-trip time---occurs and is maintained over a range of system parameters. We observe entrainment with an integer locking ratio as high as $N=20$, and propose a way to realize the effect with higher entrainment ratios and at higher finesse. This system's subharmonic response to governing equations that are invariant under discrete time translations is similar to recently proposed and reported discrete time crystals in quantum many-body systems. We discuss a route towards experimental realization of these breather-soliton time crystals, and explain how their incorporation into photonics systems could simplify proposed applications of microresonator solitons. Importantly, our work explores a new regime of microresonator Kerr-nonlinear dynamics in which the round-trip-time perturbation plays a significant role.

\end{abstract}


\maketitle

\section{Introduction}
\begin{figure}[b!]
	\includegraphics[width=8cm]{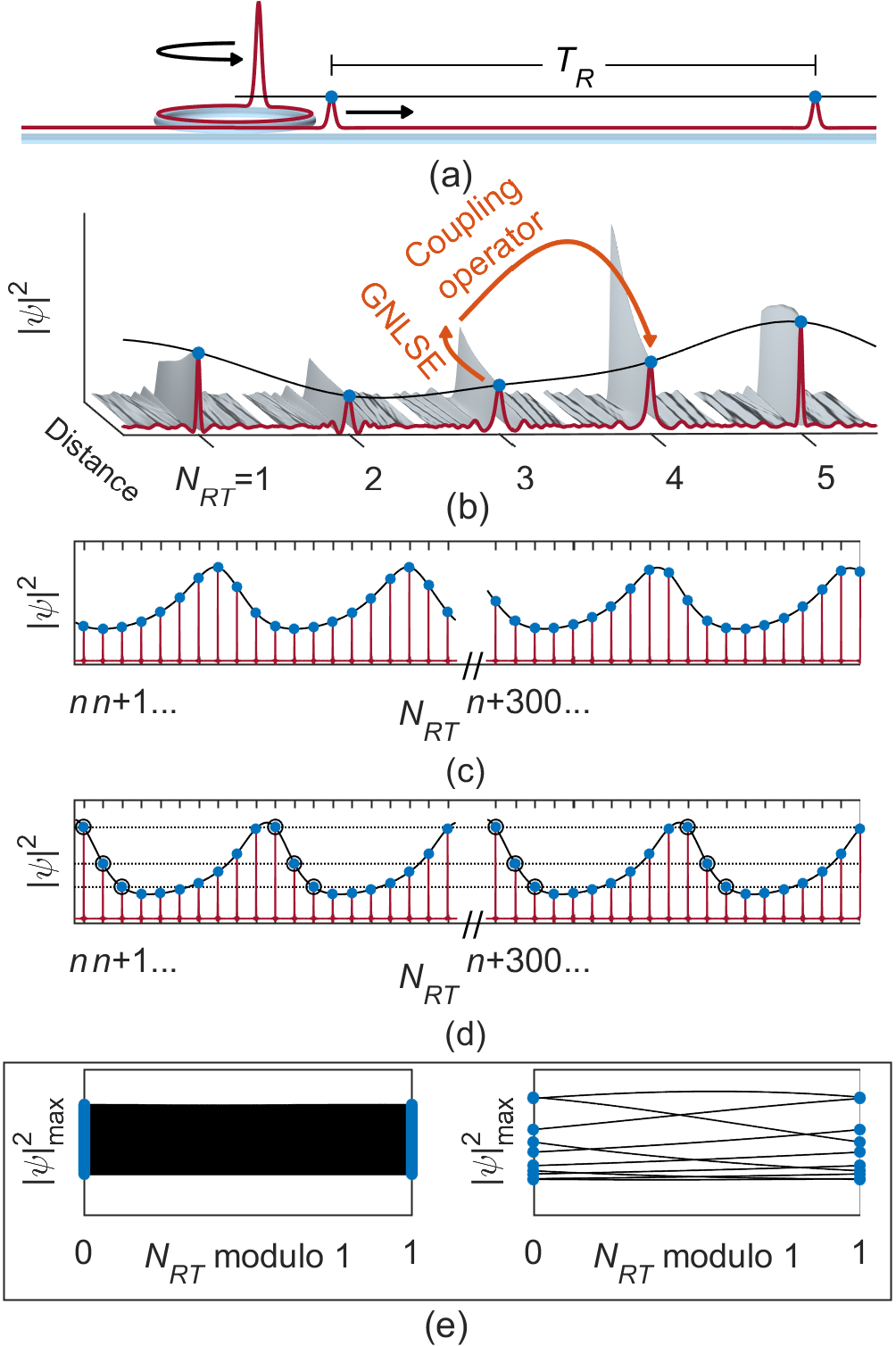}
	\caption{The physical system and the phenomenon under study. Each quantity is shown in a consistent way throughout the figure. (a) A depiction of the physical system, consisting of a Kerr-nonlinear ring resonator and a bus waveguide. The intensity profile $|\psi|^2$ of a circulating (stable, i.e. non-breathing) soliton is shown in red, along with two out-coupled pulses separated by the round-trip time $T_R$, with amplitudes indicated by blue dots. (b) A depiction of an Ikeda-map simulation of a breather soliton conducted with parameters $\alpha=4.5$, $F^2=10$, $\beta_2=-0.2$, and $\mathcal{F}=12$. The intensity of the out-coupled pulse train is shown in red as a function of round-trip times $N_{RT}$, and the gray surface plot depicts evolution (into the page) of the intensity within each round trip according to a generalized nonlinear Schrodinger equation. Blue dots mark amplitudes of the out-coupled pulses $|\psi|^2_{\mathrm{max}}$, and the black curve is a guide to the eye obtained by spline interpolation between the pulse peaks. (c, d) Depictions of breather oscillations in the unlocked (c) and locked (d, with $N=10$) cases, here for $\alpha=4.5$, $\beta_2=-0.02$, $\mathcal{F}=24.69$, and $F^2=9$ (unlocked) and $F^2=10$ (locked). In (d), we explicitly show the periodicity of the breather amplitudes for three different points in the oscillation waveform. (e) A plot of observed oscillator positions ($|\psi|^2_{\mathrm{max}}$) as a function of drive phase ($N_{RT}$ $\mathrm{modulo}$ $1$) over a simulation of $\sim8000$ round trips in the unlocked (left) and locked (right) cases. In the locked case (where $T_b/T_R=10$) the oscillator position assumes $N=10$ discrete values for each value of the drive phase; no such order exists in the unlocked case.} \label{fig:system}
\end{figure}

Coupling a continuous-wave pump laser into a passive, high quality-factor $(Q)$ Kerr-nonlinear ring resonator gives rise to rich nonlinear dynamics. In the presence of high circulating intensity enabled by high $Q$, the intensity-dependence of the refractive index due to the Kerr ($\chi^{(3)}$) nonlinearity leads to cascaded four-wave mixing, resulting in electromagnetic fields at new frequencies and the generation of broadband optical frequency-comb spectra \cite{DelHaye2007,Agrawal2007}. Of particular interest is the case where the electromagnetic fields at the new frequencies mode lock such that they periodically constructively interfere in unison, resulting in one or more localized \textit{soliton} pulses circulating the resonator \cite{Kippenberg2018,Leo2010a,Herr2014}. These solitons are able to propagate in the resonator indefinitely due to a balance between resonator dispersion and nonlinearity, and also between dissipation and the drive provided by the pump laser. If a single soliton circulates in the cavity then a pulse is out-coupled once per round trip, and a train of pulses separated by the round-trip time $T_R$ propagates away from the resonator---this is illustrated schematically in Fig. \ref{fig:system}a.

In the frequency domain, this pulse train corresponds to a coherent optical frequency comb with repetition rate $f_{\mathrm{rep}}=1/T_R$ \cite{Hall2006,Hansch2006}. Solitons generated in high-$Q$ Kerr-nonlinear microresonators (\textit{microcombs}) have been recognized as a promising candidate for realization of chip-scale optical frequency measurement and synthesis capabilities \cite{Kippenberg2018,Papp2014,Suh2016a,Marin-Palomo2017,Spencer2018,Suh2018,Trocha2018}. Much effort has been put into investigating the basic properties of these systems, and facets that have been investigated in depth include e.g. the effects of dispersion \cite{Yang2016,Brasch2016} and mode structure \cite{Savchenkov2012,Herr2014a,Yi2015,Xue2015} on the soliton spectrum, thermal effects \cite{Carmon2004,Joshi2016a,Stone2018}, spontaneous temporal ordering of co-propagating solitons \cite{Wang2017,Cole2017}, the effect of Raman scattering on the dynamics \cite{Karpov2016,Okawachi2017,Yang2017,Wang2018b}, and the soliton breathing instability.

This last effect is the spontaneous emergence of oscillations in the amplitude and temporal duration of a circulating soliton. These oscillations typically occur over many round trips with a breathing period $T_b\gg T_R$ so that they manifest as the modulation of the pulse energies in the out-coupled pulse train. This is depicted in Fig. \ref{fig:system}b and can be revealed, e.g., by impinging the pulse train on a photodetector. This phenomenon has been investigated experimentally in several works \cite{Yu2017,Guo2017,Bao2016,Lucas2017a} and is well-known outside of the field of microresonator frequency combs (see e.g. \cite{Dudley2009,Kibler2010}), but the emphasis for microcomb applications has remained on the generation of stable (non-breathing) solitons. Here we focus on breather solitons and propose a new type of spontaneous synchronization in microcombs: subharmonic entrainment of breather oscillations to the round-trip time.

It has generally been assumed that the timescale over which the intracavity field evolves in microcombs is determined by the cavity photon lifetime $\tau_{ph}$, and that the round-trip time plays little role beyond setting the repetition rate $f_{\mathrm{rep}}$ of the out-coupled pulse train. This approximation emerges in the limit of high finesse and results in the mean-field \textit{Lugiato-Lefever equation} (LLE) model of the nonlinear dynamics \cite{Herr2014,Lugiato1987,Matsko2011,Coen2013a,Chembo2013,Godey2014}. The LLE has been extremely successful in faithfully reproducing and explaining experimental results and informing system design. On the other hand, in nonlinear systems even small perturbations can lead to the emergence of qualitatively new behavior. Thus, motivated by the recent excitement regarding the proposal and observation in quantum many-body systems of discrete time crystals that violate discrete time-translation symmetry \cite{Yao2017,Choi2016,Zhang2016,Yao2018}, by the body of work regarding subharmonic entrainment of oscillations that dates back much further \cite{VanDerPol1927,Storti1988,Parlitz1997,Varangis1997}, and by previous studies of period-doubling in passive \cite{Coen1998} and lasing \cite{Soto-Crespo2004} fiber-loop cavities, we investigate the possibility of subharmonic entrainment of breather oscillations to the periodic perturbation at the round-trip time such that $T_b=NT_R$, where $N\gg1$. 

The synchronization we seek cannot be revealed within the LLE model, so in our investigations we perform simulations of individual soliton round trips. This is shown schematically in Fig. \ref{fig:system}b and described below. If this synchronization exists it would be equivalent to the spontaneous symmetry violation in quantum time crystals and to the \textit{frequency demultiplication} in an electronic system reported by van der Pol and van der Mark nearly a century ago \cite{VanDerPol1927}. As far as we know this would represent a new type of effect in microcomb nonlinear dynamics. We adopt the descriptive term \textit{breather-soliton time crystals} to emphasize the connection between our investigations and recent research outside of the field of microcombs, and to make clear the nature of discrete-time-translation-symmetry violation that we observe.

Our numerical simulations reveal subharmonic entrainment of the breather oscillations to the round-trip time in a regime of intermediate finesse; the basic effect is shown in Fig. \ref{fig:system}c-e. We focus on this regime in this initial study for two reasons---it strengthens the perturbation at the round-trip time, and it reduces the time needed to conduct simulations. We present synchronization results for $T_b/T_R=N$ up to 13. We explore the rigidity of subharmonic entrainment and find that it remains robust over a range of system parameters, and even as parameters are dynamically varied. Our simulations indicate that the strength of entrainment decays exponentially as the finesse is increased, so we discuss possible routes towards realization of the effect at higher finesse. We find that temporal sharpening of the breather oscillation peak through introduction of fourth-order resonator mode dispersion immediately allows us to extend the effect to $N=20$ without making substantial efforts at optimization. This locking ratio already indicates that the effect could be employed to electronically measure, for example, a microcomb repetition rate of 1 THz by measuring the frequency of the subharmonically-entrained breather oscillations at 50 GHz. This would allow significant simplification of proposals for integration of microcombs for applications. We conclude by discussing possibilities for experimental realization of breather-soliton time crystals.

\section{Model}

The canonical model for nonlinear dynamics in a driven, passive, Kerr-nonlinear resonator has become the nonlinear partial differential equation known as the Lugiato-Lefever equation. In normalized form, the equation reads \cite{Godey2014}:
\begin{eqnarray}
\frac{\partial\psi}{\partial\tau}=-(1+i\alpha)\psi+i|\psi|^2\psi-i\frac{\beta_2}{2}\frac{\partial^2\psi}{\partial\theta^2}+F. \label{eq:LLE}
\end{eqnarray}
Here $\psi(\theta,\tau)$ is the normalized intracavity field envelope and $F$ is the normalized pump strength (i.e. $|\psi|^2$ and $F^2$ are proportional to the circulating intensity and pump power, respectively---$F$ is assumed real without loss of generality), where normalization is taken relative to the circulating intensity and pump power at the threshold for parametric oscillation. The parameters $\alpha$ and $\beta_2$ represent the detuning of the pump laser from the nearest cavity mode and the cavity second-order dispersion, respectively, both normalized to half the cavity linewidth $\Delta\omega=1/\tau_{ph}$: $\alpha=-2(\omega_{p}-\omega_0)/\Delta\omega$ and $\beta_2=-2D_2/\Delta\omega$. Here $\omega_p$ and $\omega_0$ are the pump-laser frequency and frequency of the resonator mode closest in frequency to the pump laser, and $D_2=\frac{\partial^2\omega_\mu}{\partial\mu^2}\Bigr|_{\mu=0}$ is the derivative of the cavity free-spectral range $D_1=\frac{\partial\omega_\mu}{\partial\mu}\Bigr|_{\mu=0}$ with mode number $\mu$, evaluated at the frequency of the pumped mode (the set $\{\omega_\mu\}$ represents the cavity resonance angular frequencies, with $\mu$ the pump-referenced mode number). The field $\psi$ is defined over a co-moving azimuthal angle $\theta$ (running from $-\pi$ to $\pi$) that is analogous to the fast time in, e.g., the generalized nonlinear Schrodinger equation (GNLSE) in fiber optics, and evolves over a slow time $\tau$ normalized to the photon lifetime as $\tau=t/2\tau_{ph}$.

Analytical and numerical investigation of Eq. (\ref{eq:LLE}) and modifications thereof (e.g. to include higher-order dispersion or Raman scattering) exhibit close agreement with experimental observations. However, the LLE is not suited for revealing dynamics of the system that occur \textit{as a result of} the periodic perturbation to the system that occurs at the round trip time due to the interference of the circulating field $\psi$ with the pump field $F$ and the localized loss at the output coupler---this perturbation is assumed small and then incorporated continuously after averaging, resulting in the mean-field LLE model \cite{Haelterman1992a,Coen2013a}.

To explore round-trip-time effects one can make use of an \textit{Ikeda map} \cite{Ikeda1979,Haelterman1992a}, in which propagation of the field over the $n^{th}$ resonator round trip $\psi_n(\theta,0)\rightarrow\psi_n(\theta,L)$, where $L$ is the round-trip path length, is conducted according to the GNLSE (ubiquitous in describing propagation in Kerr-nonlinear, dispersive media \cite{Agrawal2007}), and then the field at the beginning of the next round trip is obtained by discretely incorporating out-coupling and pumping---this is depicted in Fig. \ref{fig:system}b. To define the Ikeda map relative to the LLE, we further specify the resonator finesse $\mathcal{F}=2\pi/\Delta\omega T_R$ and the coupling ratio $\eta=\Delta\omega_{ext}/\Delta\omega$, which quantifies how much of the loss is due to external coupling that occurs at the rate $\Delta\omega_{ext}$ (where $\Delta\omega=\Delta\omega_{int}+\Delta\omega_{ext}$, with internal dissipation occurring at the rate $\Delta\omega_{int}$). With these new definitions, the Ikeda map that is equivalent to the LLE in Eq. (\ref{eq:LLE}) in the limit of high finesse is \cite{Cole2018}:

\begin{eqnarray}
\psi_n(\theta,L)=G_{RT}\psi_n(\theta,0),
\end{eqnarray}
\begin{eqnarray}
\psi_{n+1}(\theta,0)=e^{-i\frac{\pi}{\mathcal{F}}\alpha}\left(1-\frac{\pi\eta}{\mathcal{F}}\right)\psi_n(\theta,L)+\frac{\pi}{\mathcal{F}}F, \label{eq:Coupling}
\end{eqnarray}
where $G_{RT}$ is an operator denoting propagation of the field $\psi$ over a single round trip, which corresponds to a distance $\Delta s=\pi/\mathcal{F}$ in a normalized spatial coordinate $s$. Propagation occurs according to a GNLSE of the form:

\begin{eqnarray}
\frac{\partial\psi}{\partial s}=-(1-\eta)\psi+i|\psi|^2\psi-i\frac{\beta_2}{2}\frac{\partial^2\psi}{\partial\theta^2}. \label{eq:NLSEn}
\end{eqnarray}

In this work we focus on the case $\eta=1$ throughout---this corresponds to the hypothetical scenario in which losses only occur through outcoupling ($\Delta\omega=\Delta\omega_{ext}$), and is approximated by a resonator that is strongly overcoupled. This is natural because we consider a regime in which the finesse is significantly lower than the value allowed by, e.g., critical coupling ($\eta=1/2$) of a resonator fabricated with high-quality materials, and setting $\eta=1$ also maximizes the strength of the periodic perturbation at the round-trip time. 

The Ikeda map incorporates the round-trip timescale into the system, which is completely absent in the LLE dynamics \footnote{To be precise, the round-trip timescale manifests in the normalization of the system parameters, e.g. the pump power, and in the size of the domain $-\pi\leq\theta\leq\pi$ relative to the dispersion coefficient $\beta_2$, which sets the duty cycle of the out-coupled pulse train.}. However, for the values of finesse that we consider here, the photon lifetime still sets the timescale for evolution of the intracavity field (see e.g. Fig. \ref{fig:Adler}a). The ratio between the photon lifetime and the round-trip time, which gives an indication of how many round trips are necessary for the field $\psi$ to change appreciably, is $\tau_{ph}/T_R=\mathcal{F}/2\pi$. 

\subsection{Determination of the breathing frequency using the Ikeda map}

In this work we employ Ikeda-map simulations to determine the oscillation frequencies of breather solitons, which allows us to identify subharmonic entrainment. Here we briefly describe the technical details of our approach.

After we have identified a point at which a breather soliton exists, the oscillation frequency is determined in the following way: An Ikeda-map simulation is initialized with a waveform $\psi_0$ that approximates the soliton solution to the LLE \cite{Herr2014} (which shares parameters $\alpha$, $\beta_2$, and $F^2$ with the Ikeda map), and then run for many round trips. Each round trip consists of evolution according to the GNLSE, Eq. (\ref{eq:NLSEn}), and then implementation of the coupling operator, Eq. (\ref{eq:Coupling}). The integration of the GNLSE is carried out using a fourth-order Runge-Kutta interaction-picture method \cite{Hult2007}, where the step size is chosen to be a (finesse-dependent, $\leq1/15$) fraction of the smaller of the nonlinearity length scale $L_{NL}$ and the dispersion length scale $L_D$ for the propagating pulse \cite{Agrawal2007}. 

The simulation duration, for example $\geq200000$ round trips for the data presented in Fig. 2, is significantly longer than a reasonable estimate of a lower bound on the required number of round trips needed to look for subharmonic entrainment. This estimate is obtained by calculating the number of round trips $N_{lock}\approx\left|1-T_{b,0}/N\right|^{-1}$ required for a breather oscillating with free-running period $T_{b,0}$ (measured in units of round-trip times) to acquire a phase of $2\pi/N$ relative to the subharmonic perturbation frequency $f_{\mathrm{rep}}/N$; this is the criterion for each possible phase relationship between the free-running breather oscillation and the periodic perturbation at the round-trip time to be sampled. An immediate conclusion is that longer simulation durations $N_{lock}$ are required to detect entrainment as the effect gets weaker, as this means that it will be observed over a narrower range $\left|1-T_{b,0}/N\right|$---note that in general one expects entrainment to be more likely when $T_{b,0}/N\approx1$.

To determine the breather frequency from the simulation data, we calculate the Fourier series for a sequence of breather amplitudes $\{|\psi_n|^2_{\mathrm{max}}\}$ of length $\geq5000$ obtained at the end of the simulation (i.e. after the system has had time to approach its long-time behavior). The breather period $T_b$ is then determined by fitting to the peak of the Fourier series near $f=1/T_b$ a function describing the Fourier series for a sine wave whose period is incommensurate with the observation time, i.e. the Fourier series for $\sin\omega t$ over the interval $0\leq t\leq T$, where $\omega T/2\pi\neq$ integer $m$.

\section{Results}

\begin{figure}[t!]
	\includegraphics[width=8cm]{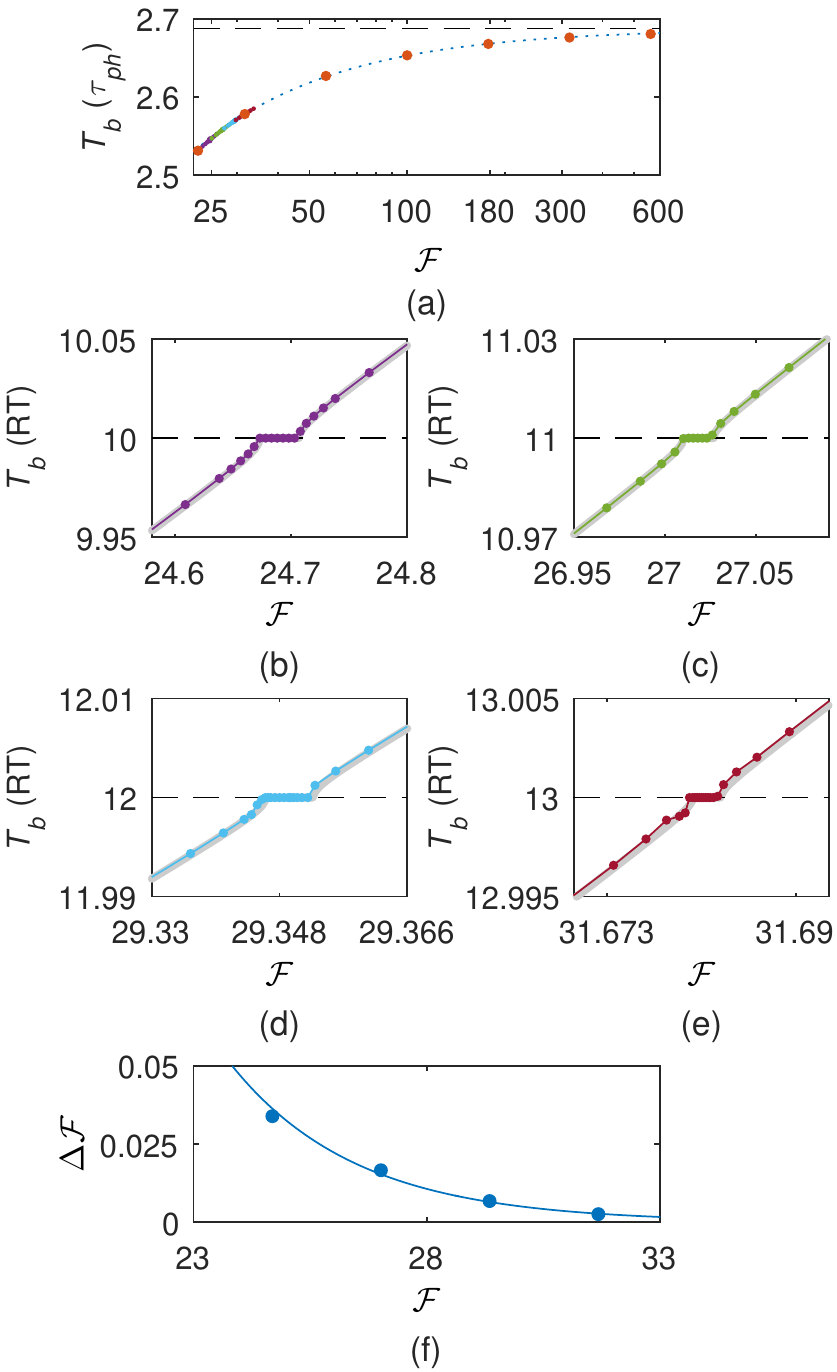}
	\caption{Exploration of breather-soliton time crystals. (a)~An investigation of the effect of varying finesse on the period of the breather oscillation, in units of photon lifetimes. The breather oscillation period in the limit of high finesse provided by the LLE is shown in dashed black. The fit shown in dotted blue is obtained from the orange dots and described in the text. The data points presented in parts b-e are also included. (b-e) Calculated breather-oscillation periods in units of round-trip times as a function of finesse over several intervals where $T_{b,0}$ is near an integer. Fits to the data using an injection locking model, described in the text, are shown in solid gray. (f) Locking range $\Delta\mathcal{F}$ as a function of finesse (dots), obtained from the injection-locking fits from parts b-e, with a decaying exponential fit to the data indicated by the solid line. } \label{fig:Adler}
\end{figure}

Here we present our numerical simulations of breather-soliton time crystals. First we explore the effect of varying the resonator finesse  on the breather oscillations (with fixed pump power and detuning), and we find intervals of finesse over which $T_b$ locks to an integer $N$, resulting in an entrainment plateau (Fig. \ref{fig:Adler}). Then we explore these entrainment plateaus as a function of pump power $F^2$ and detuning $\alpha$; these parameters are readily tuned in experiment, which provides a route for accessing entrainment. Finally, we investigate the rigidity of a breather-soliton time crystal in the presence of dynamical variations of the pump power. 

We begin by examining the effect on breather oscillations of departing from the high-finesse limit. For the remainder of the paper we focus on a breather soliton at or near the point $\alpha=4.5$, $F^2=10$, $\beta_2=-0.02$, and $\eta=1$. In Fig. \ref{fig:Adler}a we show the period of breather oscillations as a function of the resonator finesse at this point; the breather period relative to the photon lifetime changes little as the finesse is varied. We find that the departure of the breather period from the high-finesse limit, where the breather period $T_{b,\tau_{ph}}$ is determined within the LLE model, is well described by $T_{b,\tau_{ph}}(\mathcal{F})=T_{b,\tau_{ph},LLE}-A/\mathcal{F}$, where $A\sim3.47$ (here and below we use the subscripts `$\tau_{ph}$' and `$RT$' to make explicit the units of measurement for $T_b$; otherwise it should be assumed that the units are round-trip times). From preliminary investigations of other points $(\alpha,F^2,\beta_2,\eta)$, it appears that this tidy form of the fit may not be universal---in particular, other powers of $\mathcal{F}$ may be required for agreement with the observations.

Even though the breather period changes little in units of photon lifetimes over the range $25\leq\mathcal{F}\leq600$, it varies greatly in units of round-trip times through the relation $T_{b,RT}=T_{b,\tau_{ph}}\cdot\mathcal{F} /2\pi$---thus as the finesse is varied the breather period $T_b$ traverses many integer values. In Figs. \ref{fig:Adler}b-e we present calculations of the breather period for intervals of finesse over which $T_b$ is close to an integer. Clearly visible in this data are plateaus of subharmonic entrainment, where $T_b$ deviates from the expected value and rigidly snaps to an integer.

In Figs. \ref{fig:Adler}b-e we also present fits of the entrainment data to the Adler injection locking model \cite{Adler1973}. This model provides a prediction for the difference between the breather frequency $f_b=1/T_b$ and the nearby subharmonic perturbation frequency $f_{\mathrm{rep}}/N$ in terms of the free-running breather frequency $f_{b,0}(\mathcal{F})=1/T_{b,0}(\mathcal{F})$ and the locking bandwidth (in frequency) $\delta$ that determines the locking range: 
\begin{eqnarray}
f_b=f_{\mathrm{rep}}/N+\mathrm{Sign}(\Delta)\times\mathrm{Re}\,\sqrt{\Delta^2-\delta^2}, \label{eq:Adler}
\end{eqnarray}
where $\Delta=f_{b,0}(\mathcal{F})-f_{\mathrm{rep}}/N$ and $\delta$ is determined by fitting the data. For Figs. \ref{fig:Adler}b-e $f_{b,0}(\mathcal{F})$ is assumed to be locally linear and is obtained by fitting data far from the entrainment plateau. From the calculated values of $\delta$ we can determine the locking range $\Delta\mathcal{F}$ in each case, and we plot these values as a function of the finesse in Fig. \ref{fig:Adler}f. We observe that the locking range is fit well by a decaying exponential. This rapid decrease in the locking range can be expected for several reasons: First, the magnitude of the periodic perturbation at the round-trip time decreases with increasing finesse. Second, we expect the sensitivity of the system to shifts in the timing of out-coupling relative to a hypothetical breather oscillation envelope (represented conceptually by the black curves in Fig. \ref{fig:system}) to be related at least qualitatively to $\partial A_b/\partial N_{RT}$, where $A_b$ is the amplitude of the breather envelope and $N_{RT}$ is the slow time in units of round trips. This sensitivity will decrease as $T_b$ increases with increasing finesse.

\begin{figure}[t!]
	\includegraphics[width=8cm]{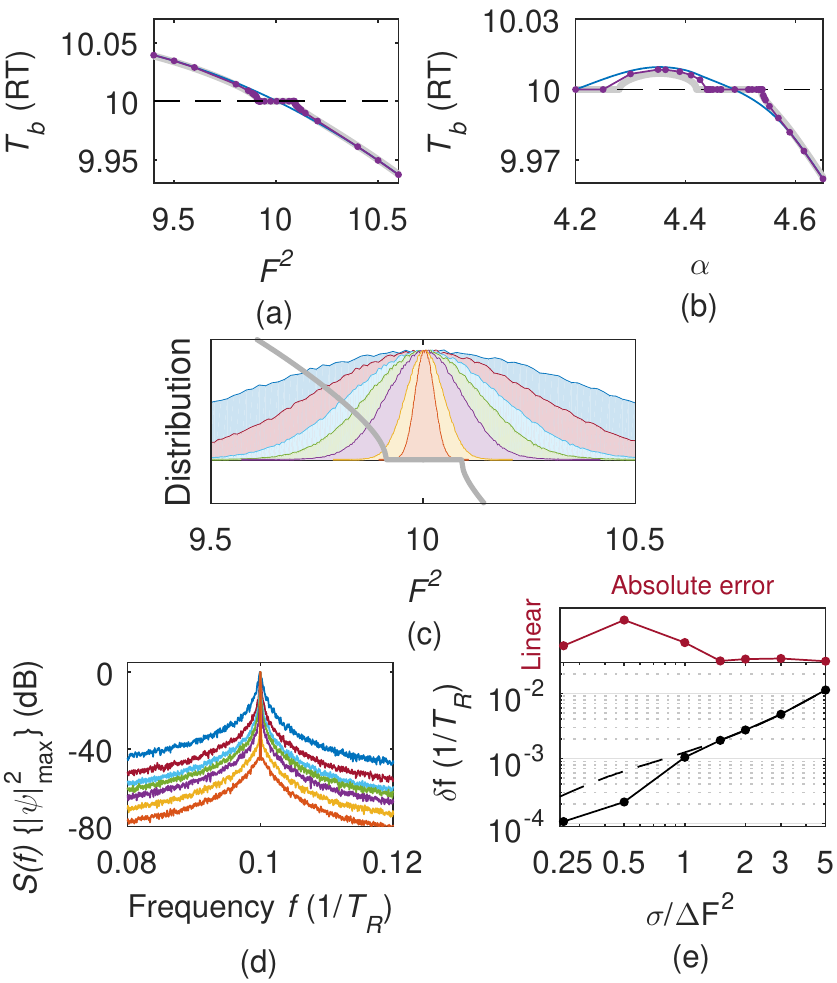}
	\caption{Rigidity of subharmonic entrainment with $N=10$ ($\mathcal{F}=24.69$) in the presence of variation of system parameters $F^2$ and $\alpha$. (a) Entrainment plateau for variation of $F^2$, with approximation of $T_{b,0}(F^2)$ (blue) and fit to the injection-locking model (gray) obtained as described in the text. (b)~Two entrainment plateaus for variation of $\alpha$. This observation underscores the opportunity for more in-depth investigations of breather dynamics in future work. (c-e) Rigidity of entrainment when $F^2$ is dynamically varied using additive white Gaussian noise. (c) Distribution of $F^2$ values for each of seven simulations, with the fit from part a for scale (gray).  (d) Calculated spectra of out-coupled breather amplitudes for each distribution after observation over the second half of a simulation of $2^{19}$ round trips. Colors correspond to c. The spectra are scaled to the same peak amplitude; spectra obtained for narrower $F^2$ distributions have a higher-amplitude coherent spike at $f=0.1/T_R$ (i.e. spectral wings are lower). (e) Thirty-dB width $\delta f$ of the spectral peak at $f=0.1/T_R$, calculated with a resolution bandwidth of $\sim1\times10^{-4}/T_R$. The $x$-axis indicates the ratio of the standard deviation $\sigma$ of the normally-distributed $F^2$ values to the locking range $\Delta F^2$. The dashed line is a model for the results in the absence of subharmonic entrainment as described in the text, and indicates that a qualitative change occurs due to the influence of entrainment as $\sigma$ is decreased. The difference between the simulation results (black dots) and the model (dashed line) is indicated on a linear scale above the plot.} \label{fig:rigidity}
\end{figure}

Next, we investigate the entrainment plateaus as a function of the system parameters $F^2$ and $\alpha$, which represent the pump power and laser-cavity detuning, respectively, and present the results in Fig. \ref{fig:rigidity}. Fig. \ref{fig:rigidity}a shows the entrainment plateau as a function of $F^2$, and Fig. \ref{fig:rigidity}b shows two entrainment plateaus as a function of $\alpha$. As in Fig. \ref{fig:Adler}, we present fits to the data using Eq. (\ref{eq:Adler}). To implement the injection-locking fit in Fig. \ref{fig:rigidity}a, we use spline interpolation between data points far from the injection locking plateau to approximate $f_{b,0}$, which allows us to carry out the fit without knowing the precise form of the dependence of $f_{b,0}$ on $F^2$. For Fig. \ref{fig:rigidity}b, because we lack data on the free-running oscillator frequency $f_{b,0}(\alpha)$ for $\alpha$ below the plateau centered near $\alpha=4.5$ (as all this data is strongly pulled), we use a two-step model in which the data above this plateau is used to generate an initial fit, and then this fit is used to approximate the free-running frequencies $f_{b,0}(\alpha)$ over the full $\alpha$-interval. The fits (gray and blue curves) presented in Fig. \ref{fig:rigidity} should be regarded as qualitative approximations.

We also investigate breather oscillations in the presence of dynamical variations of the pump power around a point at which subharmonic entrainment occurs. We calculate the spectrum of the breather amplitudes $|\psi|^2_{\mathrm{max}}$ in the presence of time-varying $F^2$ values drawn from normal distributions with standard deviations $\sigma$ up to $5\Delta F^2$, where $\Delta F^2=0.087$ is the locking range obtained from Fig. \ref{fig:rigidity}a (half the width of the plateau); this value $5\Delta F^2$ corresponds to $\sim4$ $\%$ of the mean value $F^2_0=10.0034$. Figs. \ref{fig:rigidity}c-e summarize the results of this study. We highlight Fig. \ref{fig:rigidity}e, which presents a calculation of the thirty-dB linewidth of the breathing frequency tone as a function of $\sigma$. Included in this plot is a phenomenological model for the data in which the tone is assumed to be a Lorentzian that is sampled with the appropriate resolution bandwidth and that has linewidth proportional to $\sigma^{1.8}$. One arrives at a model in which the linewidth is proportional to $\sigma^2$ by assuming that the instantaneous breathing frequency varies linearly with $F^2$, noting then that the power spectrum of fluctuations in the breather frequency is proportional to $\sigma^2$, and using the fact that in the case of white noise the linewidth is proportional to this power spectrum \cite{DiDomenico2010}; the phenomenological dependence $\sigma^{1.8}$ provides a better fit to the data at high $\sigma$, likely due to the fact that the breathing frequency does not actually vary linearly with $F^2$. This model agrees well with the data at high $\sigma$, but at low $\sigma$ the data drops below the model; this indicates that subharmonic entrainment exerts a stabilizing influence on the breathing frequency as $\sigma$ is decreased.

\section{Discussion}
We have presented evidence for subharmonic entrainment of breather-soliton oscillations to the periodic perturbation occurring at the round-trip time in passive, driven, Kerr-nonlinear ring resonators. This effect violates discrete time-translation symmetry in the same manner as the discrete quantum time crystals that have been recently demonstrated.

Our prediction of breather-soliton time crystals in microcombs goes beyond what has been investigated experimentally and what can be explored within the canonical mean-field LLE model for microcomb nonlinear dynamics. We expect this proposal to open up a new avenue for research in this field, as effects like this one related to the round-trip time have not yet been investigated. This new phenomenon could prove useful in tailoring microcombs for applications.

An obvious limitation of this preliminary study is that we have investigated a regime of intermediate finesse, with values of finesse between 24 and 32 explicitly considered. The finesse of microresonators that are employed in experiments is typically in excess of 1000 (sometimes greatly so), and lowering the finesse comes with the drawback of increasing the threshold power for comb generation. Moreover, we have seen that the strength of the subharmonic entrainment effect, as indicated by the locking range $\Delta\mathcal{F}$, falls of exponentially with increasing finesse. However, there are many tools that may aid in overcoming this apparent obstacle that could be investigated in future work. First, we have only investigated a single LLE point (defined by $\alpha$, $\beta_2$, and $F^2$) in depth---this leaves the remainder of parameter space open for exploration. Moreover, there are effects that have been investigated within the context of the LLE that we have not investigated here. These include, for example, Raman scattering and other higher order nonlinearities, and also higher-order terms in the Taylor expansion of the mode frequencies (which we have truncated at second order, $\beta_2=-2 D_2/\Delta\omega$). Additionally, one could consider azimuthally varying the nonlinear, dissipative, or dispersive properties of the resonator to increase the strength of round-trip-time effects (e.g. \cite{Bao2015}). Finally, the coupling rate $\Delta\omega_{ext}$ could be spectrally varied so that the round-trip-time perturbation at the pump frequency is strong while out-coupling of the other comb modes is weaker, which could help to mitigate the increases in threshold power that come with lower finesse.

\begin{figure}[b]
	\includegraphics[width=8cm]{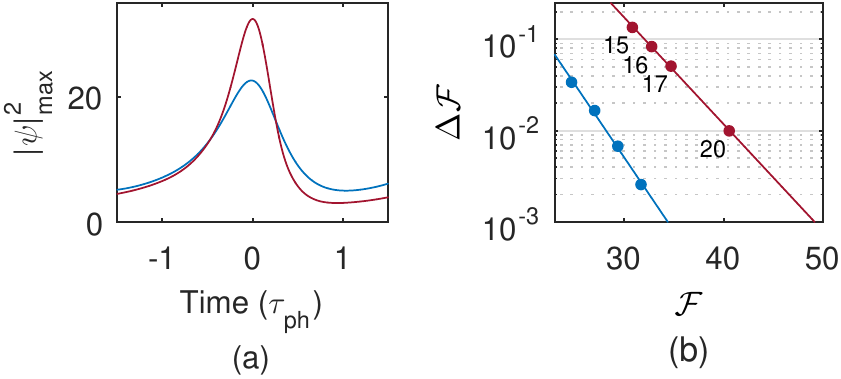}
	\caption{Extension of subharmonic entrainment to higher finesse. (a) LLE simulations of a breather soliton's amplitude at the point $\alpha=4.5$, $F^2=10$, $\beta_2=-0.02$, for the cases of $\beta_4=8\times10^{-6}$ (red) and $\beta_4=0$ (blue). The inclusion of fourth-order dispersion with sign opposite that of $\beta_2$ sharpens the breather oscillation, making it more sensitive to changes in the timing of the out-coupling of pulses. (b) Locking range $\Delta\mathcal{F}$ observed in Ikeda-map simulations of the two breathers from part a. The case of $\beta_4=0$ was analyzed extensively above; including $\beta_4=8\times10^{-6}$ both shifts the locking-range curve up and reduces its slope. Locking ratios $N$ for the case $\beta_4\neq0$ are indicated next to the data points.} \label{fig:extension}
\end{figure}

Intuitively, one can expect that sharpening the breather oscillation waveform to increase the sensitivity $\partial A_b/\partial N_{RT}$ will facilitate realization of this effect at higher finesse. To test this hypothesis, we incorporate fourth-order dispersion into the LLE and the GNLSE for the Ikeda map (Eqs. (\ref{eq:LLE}) and (\ref{eq:NLSEn}), respectively) by adding the term $+i\frac{\beta_4}{4!}\frac{\partial^4\psi}{\partial\theta^4}$ to the right-hand sides of these equations, where $\beta_4=-2 D_4/\Delta\omega$, $D_4=\frac{\partial^4\omega_\mu}{\partial\mu^4}\Bigr|_{\mu=0}$. Upon setting $\beta_2=-0.02$ and $\beta_4=8\times10^{-6}$, we find that an LLE simulation of the breather oscillations reveals sharper and higher-amplitude oscillations than in the case of only second-order dispersion with $\beta_2=-0.02$. Intuitively this can be understood by noting that $\beta_4$ with appropriate magnitude and with sign opposite that of $\beta_2$ reduces the integrated dispersion $D_{int}(\mu)=\omega_\mu-(\omega_0+D_1\mu)$, which quantifies the dependence of the local comb-resonator detuning on mode number $\mu$, far from the pump \cite{Brasch2016}. This facilitates broad bandwidth, temporal pulse compression, and high peak power, but only for a pulse that already has sufficient bandwidth to sample this region of reduced integrated dispersion---therefore temporal narrowing and increased peak power occurs for pulses that are already temporally short, i.e. when the breather is near its peak amplitude. We explore subharmonic entrainment of these oscillations using the Ikeda map model, and find greater than ten-fold improvement in the locking range as a function of finesse. These results are summarized in Fig. \ref{fig:extension}. The improved locking range allows us to observe the effect up to at least a locking ratio of $N=20$.

This increased locking ratio immediately suggests an application for breather-soliton time crystals. Presently, promising proposals for integration of microcombs into low size, weight, and power photonics packages make use of a comb with an extremely high ($\sim1$ THz) repetition rate, which facilitates achievement of the octave-spanning spectrum required for full stabilization of the system \cite{Spencer2018,Briles2018}. However, a second lower-repetition-rate comb must be incorporated into the system to enable measurement of the first comb's repetition rate, which is outside the bandwidth of electronic measurement systems. The possibility of generating a breather soliton in the $\sim1$ THz resonator and measuring the frequency of entrained breather-soliton oscillations as a proxy for the repetition rate could therefore greatly simplify this system. Indeed, the locking ratio of $N=20$ we observe in the presence of fourth-order dispersion is already high enough to enable indirect measurement of a $1$ THz repetition rate through measurement of entrained breather oscillations at 50 GHz.

At present it is not clear what kind of resonator design will best facilitate realization of breather-soliton time crystals in high-finesse microresonators. However, this phenomenon could be immediately investigated experimentally using fiber-loop `macro-ring' resonators. The dynamics in a passive, driven loop of nonlinear fiber are formally equivalent to those of a microcomb, but fiber loops typically have finesse on the order of $\sim100$ instead of $\sim1000$. Many facets of Kerr-nonlinear dynamics have been investigated in fiber loops (e.g. \cite{Leo2010a,Jang2015,Wang2017,Wang2018b}), and this is certainly a promising route for experimental realization of the breather-soliton time crystals we have proposed here. We present our results with the hope that they will excite interest in this fundamentally new type of behavior, spur a new course of microcomb research, and prove useful in developing this technology towards maturity for applications.

\begin{acknowledgments}
This work was funded by the DARPA DRINQS and DODOS programs and NIST. We thank Abijith Kowligy and Liron Stern for comments on the manuscript.
\end{acknowledgments}

\bibliography{BreatherTimeCrystals_v1}

\end{document}